\documentstyle[aps,prl,twocolumn]{revtex}
\begin{document}
\draft
\title{Theoretical evidence for efficient $p$-type doping of GaN 
using beryllium}
\author{ Fabio Bernardini and Vincenzo Fiorentini}
\address{INFM-Dipartimento di Scienze Fisiche,
Universit\`a di Cagliari, via Ospedale 72, I-09124 Cagliari, Italy}
\author{Andrea Bosin}
\address{TECHSO S.p.A., Elmas (CA), Italy}

\date{\today}
\maketitle 
\begin{abstract}
Ab initio  calculations predict that
 Be is a shallow acceptor in GaN. Its thermal ionization 
energy is 0.06 eV in wurtzite GaN; the level 
 is valence resonant in the zincblende phase.
Be incorporation is severely limited by
the formation of  Be$_3$N$_2$. We show however that  co-incorporation
with reactive species can   enhance the   solubility.
H-assisted incorporation should lead to high doping 
levels in MOCVD growth after  
post-growth annealing at about 850 K. Be-O 
co-incorporation 
 produces high Be and O concentrations  at MBE
growth  temperatures.
\end{abstract}

\noindent
Gallium nitride is a base material for 
green-to-UV optoelectronics, and piezoelectric and 
high-power devices. For these applications, controllable doping  is an obvious
 necessity. Si$_{\rm Ga}$  is a suitable 
shallow donor, but the quest for the optimal acceptor still continues.
The current-best $p$-dopant\cite{noi} is Mg$_{\rm Ga}$,
with which hole concentrations
in the 10$^{17}$ cm$^{-3}$ range
have been achieved  at room temperature
(despite its ionization energy of $\sim 0.2$ eV),
and whose activation mechanism is still
under scrutiny \cite{neuapl}.

We show here that substantial improvements can be achieved
by doping GaN with Be:   ab
initio calculations of formation energies,
impurity levels, and atomic geometries of Be in GaN show 
that Be is the shallowest acceptors reported so 
far in GaN.
Calculated  solubilities  indicate barely acceptable doping 
levels in N--rich growth conditions at high growth 
temperatures, while complete
compensation occurs  in Ga--rich 
conditions due to  N vacancies. We 
 show however that the co-incorporation of
Be with H or O strongly enhances its  solubility
(otherwise limited by the formation of Be nitride) and
may  lead to
$p$-type doping.

\paragraph*{Method -- }
By means of density-functional-theory (DFT) \cite{dft} 
ultrasoft-pseudopotential plane-wave calculations of
total energies and forces in  doped GaN (details are  
reported elsewhere \cite{noi,noi2,mrs}), 
we predict from first principles the formation 
energies and thermal ionization energies,
and ensuing  
dopant and carrier concentrations of Be in GaN.
The carriers concentration  at temperature T
due to a single acceptor  with thermal
ionization energy $\epsilon[0/-]$ is 
\begin{equation}
n = D\ {\rm exp}\, (- \epsilon [0/-]/{\rm k_B T}) \equiv D\, \cal{E}.
\label{carrier}
\end{equation}
The impurity concentration $D$ in  thermal equilibrium is
\begin{equation}
D = N_s\ {\rm exp}\, (-E_{\rm form}/{\rm k_B  T_g})
\label{cul}
\end{equation}
for a  growth temperature of T$_{\rm g}$, with $N_s$=
4.33$\times$10$^{22}$ cm$^{-3}$  available sites,
and a  formation energy   $E_{\rm form}$.
Omission of the entropic factor exp (S$_f$/${\rm k_B}$)
(formation entropies are currently impossible to calculate
reliably)  does not alter our conclusions.
The formation energy for Be$_{\rm Ga}$ in charge state {\it Q} is 
\begin{equation}
E_{\rm form} (Q) = 
E^{\rm tot}(Q)  - \sum_X n^{\rm X} \mu^{\rm X} + Q (\mu_e + E_v^Q),
\label{eform}
\end{equation}
where  $\mu_e$ is the electron chemical potential,
$E^{\rm tot}(Q)$ is the total energy of the defected supercell  
in  charge state $Q$, $E_v^Q$ its top valence band energy,
$n^{\rm X}$ and  $\mu^{\rm X}$
the number of atoms of the involved species (X=Ga, N, Be)
and their chemical potentials.
The latter potentials  must satisfy the 
equilibrium conditions with GaN and Be-X compounds. To obtain 
maximum solubility, we will always be choosing
 the highest  $\mu^{\rm Be}$ 
compatible with the relevant  solubility limit. Then,
a single chemical potential (e.g. $\mu^{\rm N}$) remains to be 
determined by the imposed (N--rich $\leftrightarrow$  Ga--rich) growth 
conditions. Since T=0 in our calculations, $\mu_e$ is synonimous with 
the Fermi energy.

The thermal ionization level $\epsilon[0/-]$ of a single acceptor
is by definition the formation-energy difference of the two 
charge states $Q$=0 and $Q$=--1 at $\mu_e$=0. Thermal ionization
corresponds to hole release from (i.e., electron  promotion into) the 
acceptor extrinsic state with concurrent geometric relaxation. 
Optical  transition energies can be obtained as e.g.
$E_{\rm gap} - \epsilon[0/-] - E_{\rm FC}$ 
for conduction band--to--acceptor recombination. 
$E_{\rm FC}$  is
 the Franck-Condon shift for the $Q$=--1 state in the 
equilibrium geometry of the $Q=0$ state. For Be,
 E$_{\rm FC} = 0.04$ eV \cite{mrs}.

The above quantities are DFT total energy differences
and as such  exact in principle, were it not for the
local density approximation\cite{dft} and other technicalities. 
Among the latter, the  one introducing
the largest uncertainty is the alignment  of
the energy  zeroes for the different charge states. 
The resulting error bar 
is typically $\pm$ 0.1 eV in the ionization energies, but 
in fact lower ($\sim$ 0.05 eV) for Be$_{\rm Ga}$ (see 
Refs. \cite{noi,noi2} for details).

\paragraph*{Results -- }
The thermal level of Be$_{\rm Ga}$ is 0.06 eV in wurtzite GaN.
This value agrees well with our previous estimate \cite{mrs} based on 
a different (and less accurate)   technique. 
In the zincblende phase, the level is found to 
be resonant with the valence band. 
The difference   between
 zincblende and wurtzite is due to
symmetry-related    electronic structure
features, namely  the impurity state being  mostly 
localized on {\it all} the four
impurity-neighboring N atoms in zincblende, while only on the
 $a$-plane neighbors in wurtzite (see Ref. \cite{noi2} for more).
Despite this minor deviation, the overall  energetics
is essentially the same in the two phases (the numbers 
below are for wurtzite).

Be$_{\rm Ga}$  is a shallow dopant in GaN,
with an activation efficiency      $\cal{E} \sim$ 0.1
(see Eq.\ref{carrier}) at 
room temperature.
To evaluate its chemical  concentration, we calculated its formation
energy as a function of  growth conditions and of the
 Fermi energy (accounting for the  dependence of the latter on  charge 
neutrality conditions \cite{vdw} involving background doping and
competing defects).
For neutral Be, $E_{\rm form}$ is 2.29 eV (2.83 eV)  in   N--rich (Ga--rich) conditions;
for  $\mu_e > \epsilon[0/-]$,
 the negative state is favored
($E_{\rm form}$=2.35 eV -- $\mu_e$).
These high formation energies stem from the strict solubility
limit imposed by the formation of Be$_3$N$_2$ (whose total 
energy is also calculated ab initio).
Therefore, the Be concentration achievable in 
GaN is typically quite low.
Also,  competing  defect configurations  exist: we considered
 the nitrogen vacancy V$_{\rm N}$, two kinds of 
interstitial Be (in the wurtzite channels and in
the ``trigonal'' cage), and 
 heteroantisite Be$_{\rm N}$ (the latter
being  ruled out by a formation energy of $\sim$ 8 eV).

The most favorable growth condition  for Be$_{\rm Ga}$
incorporation is {\it N--rich}.
 V$_{\rm N}^+$ is an ionized single donor with a formation energy  of
2.32 + $\mu_e$
(possibly behaving as multiple donor in the extreme $p$ limit
\cite{neuapl}). 
The Be interstitials are double donors over most of the relevant 
Fermi level range (since $\epsilon[2+/+] \sim 2.3 $ eV),
with   formation energies of 
2.39 eV + 2 $\mu_e$ (cage) and 2.43 eV
+ 2 $\mu_e$ (channel). Therefore, none of these defects is 
ever competitive with  
 Be$^{0}_{\rm Ga}$ ($E_{\rm form}$=2.29 eV) or 
 Be$^{-}_{\rm Ga}$ ($E_{\rm form}$=2.35 eV -- $\mu_e$).
In particular, compensation  by V$_{\rm N}^+$ is suppressed.  
However,
at typical MOCVD growth temperatures of 1500 K, the
 maximum concentrations reached
is only  $D \simeq$ 10$^{17}$ -- 10$^{18}$/cm$^{3}$ assuming 
formation entropies between zero and 5 k$_{\rm B}$
(the entropic factor enhances the solubility).
In wurtzite GaN at room temperature,
thanks to Be's  shallowness, this implies
 a doping level of 10$^{16}$--10$^{17}$ cm$^{-3}$.
For comparison,  the calculated concentration \cite{noi2}
of Mg$_{\rm Ga}$ in the same conditions is
 about 10$^{19}$/cm$^{3}$ ($E_{\rm form}$=1.40 eV),
but the carrier concentration is low because of
$\cal{E} \sim$ 10$^{-4}$  (Eq.\ref{carrier}) at room 
temperature, as implied by the calculated ionization energy 
of 0.23 eV.
 
In {\it Ga--rich} conditions the situation is more critical. The
 formation energy of V$_{\rm N}^+$ is 0.67 eV + $\mu_e$, 
whereas that of Be$_{\rm Ga}^-$ is  2.89 -- $\mu_e$.
This results in complete compensation, with the
Fermi level pinned at about 1.1 eV above the valence band top.
The interstitials, again double donors
with  formation energies of 1.29 eV + 2 $\mu_e$ (cage)
and 1.33 eV + 2 $\mu_e$ (channel),  are competitive 
with Be$_{\rm Ga}$ only in $p$-type conditions 
(Fermi levels below $\sim$ 0.4 eV). Their action is anyway preempted by 
 V$_{\rm N}^+$, which pins the Fermi level much higher in 
the gap.

\paragraph*{Co-incorporation -- } 

As shown earlier, Be {\it alone} could never 
be incorporated massively in GaN. An abundant supply of  Be
in the growth process would eventually cause  Be$_3$N$_2$ to form, with
detrimental effects for GaN crystal quality \cite{vari}.
A possible  way to  enhance the
 solubility of Be in GaN is  ``incorporation catalysis'' by reactive 
species. 

We consider first the  H-assisted incorporation
suggested recently \cite{neuapl} for Mg. The
formation energy of a (neutral) Be-H complex in its equilibrium 
configuration in GaN is
0.32 (0.87) eV/complex in N--rich (Ga--rich) conditions (solubility 
limits: Be$_3$N$_2$ and H$_2$ molecules):  the concurrent
 incorporation of Be and H in GaN is therefore efficient. 
Accounting for competing species and charge neutrality,
we find that as much as 10$^{19}$ to 10$^{21}$  Be's and 
H's per cm$^{3}$ get  incorporated in the Ga--rich and N--rich limits
respectively,  at a growth temperature of  1500 K. 

However, the  acceptor is compensated by H, which must 
be  eliminated from the sample  to activate
$p$-type conduction. 
 After H is detached from the complex, Be is still  compensated, since H is 
in its positive charge state \cite{mrs,neuprl}.  H should thus be
removed from the $p$-doped  region by annealing in e.g. a N$_2$
atmosphere.  Being quite stable, the Be-H complex  
is easy to incorporate but difficult to break 
up: the minimum binding energy difference  between Be-H and 
isolated Be$^{-}_{\rm Ga}$ plus isolated H$^+$ in its equilibrium
configuration (N backbonding site) is 1.8 eV \cite{mrs,noi3}. We
estimate the
barrier for detachment to be at least of order 2.1 eV.
Within a simple desorption-like description,
 the  annealing temperature T$_{\rm ann}$  needed to break up 
H-Be complexes may be estimated from  the rate 
$$R \simeq \nu\, {\rm exp}\, (-E_{\rm barr}/{\rm k_B T_{\rm ann}}).$$
The relevant attempt frequency $\nu$ is that of  the  H-Be bond center
wagging mode, since the complex ground  state is H 
at a Be-N bond center \cite{mrs}. We have previously calculated \cite{mrs} 
the stretching frequency for the H-Mg complex to be about 
3000 cm$^{-1}$ in the equilibrium  N-backbonding site; as  this Mg-H
frequency essentially originates from N-H pairing \cite{mrs,neuprl}, it is
 expected to be similar for H-Be. As wagging frequencies of H  at bond
centers are usually about 10 times lower than stretching 
frequencies  \cite{sibc},  we pick  
$\nu$=100 THz $\sim$ 300 cm$^{-1}$. With these parameters, massive 
break-up of H-Be sets in at about 850 K, which  seems 
a practicable post-processing temperature.

 Be-H co-incorporation seems possible.
The  analogous idea  of reactive  co-doping \cite{bra}
also appears  to be  useful. In recent experiments \cite{bra},
Be and O concentrations in the 5$\times$10$^{20}$ cm$^{-3}$ range
have been reported in zincblende GaN grown by MBE at 950 K in
moderately  N--rich and oxygen-contaminated conditions.
Importantly, experiment  \cite{bra} indicates that    Be
 prevails over O by about 20 \% ($D$[Be]$=5 \times 10^{20}$ cm$^{-3}$,
$D$[O]$=4 \times 10^{20}$ cm$^{-3}$), leading to $p$--type doping.

We find that the  co-incorporation of Be and O  favors
the precipitation of BeO
over the
formation of the individual solubility-limiting compounds
 Be$_3$N$_2$ and
 Ga$_2$O$_3$, because the  reaction 
\begin{equation}
{\rm Be_3 N_2 + Ga_2 O_3 \rightarrow 2\, GaN
+ 3\, BeO}
\label{beoeq}
\end{equation}
is   strongly  exothermic (6.4 eV). Therefore,
Be and O can be  incorporated {\it concurrently} in large amounts into
GaN. However,  the incorporation of Be {\it in excess of} O (as it appears
to occur
 in the co-doping experiments \cite{bra}) is still affected
 by the  Be$_3$N$_2$
solubility limit, 
 so that  the net Be concentration cannot be larger than that
 obtained in the absence of O. 
In consideration of the above-cited
 experimental evidence by  Brandt {\it et al.} \cite{bra}, 
our results, obtained assuming the validity of
equilibrium thermodynamics,  imply that 
non-equilibrium effects  are operative in the 
actual co-doping process. Such as-yet unknown  effects may be 
 related to  growth kinetics, as is quite plausible 
considering  the highly non-equilibrium nature of the MBE growth process. 

One may speculate
 that  the presence of high chemical concentrations 
of Be   (albeit
 O-compensated) will  make it more likely for small 
non-equilibrium
 concentration fluctuations to occur in an interesting range
 (say $\sim$10$^{19}$ cm$^{-3}$, i.e. $\sim$ 5 \% of 
the total concentration), whatever 
the exact cause of
such fluctuations. This hypothesis is  expecially
intriguing, since the same mechanism may apply to III-V nitrides
 in general: we found in fact that  exothermic reactions
analogous to Eq. (\ref{beoeq})  hold for several Group-II dopants
(Be, Zn, Mg, Ca) in AlN and InN as well as in GaN, so that O-acceptor
co-incorporation will occur efficiently in most cases. A treatment of 
this issue is beyond the scope of  this work, and is deferred to a future
work.

We note that calculations show also that a neutral BeO
pair substituting for a GaN pair is bound by 1.0 eV/pair with respect 
to over isolated  Be$_{\rm Ga}^-$ and O$_{\rm N}^+$: 
this  path for  O--Be donor-acceptor
compensation may thus be
 preferred over hole-electron recombination.    BeO pairs
 carry at most a dipole, which scatters carriers less
efficiently  than  a charged center. 
An enhanced carrier mobility
should ensue, in qualitative agreement with unusually
 high mobilities being observed 
in experiment \cite{bra}.

In summary,  ab initio calculations predict that Be is 
a shallow acceptor in GaN.
Be solubility in GaN is low due to Be$_3$N$_2$
formation, so that  hole concentrations at room temperature
of only  $\sim$10$^{17}$ cm$^{-3}$
are obtained from  N--rich, high-temperature  growth.
 In Ga--rich conditions, vacancy-acceptor compensation 
produces semi-insulating  material. The
presence of H  enhances Be solubility  in GaN, although acceptor 
activation requires annealing at about 850 K. 
Our finding of  massive co-incorporation of 
 O and Be in GaN grown in an O-contaminated atmosphere confirms
in part the results of recent experiments 
\cite{bra}. In conclusion, further consideration of
Be in actual growth  experiments may open the way to a more effective
$p$-doping  of  GaN.

This work was supported in part by the European Union
through Contract BRE2-CT93-0526,  by CINECA Bologna through
Supercomputing  Grants 93-1-102-9 and 95-1506, and by CRS4, Cagliari, Italy,
within an agreement with Cagliari University.
We thank D. Vanderbilt and R. Valente for their
codes, O. Brandt 
for useful discussions and for making Ref. \cite{bra}
available prior to publication, and 
 R. Magerle and  H. Riechert  for useful discussions.


\begin{references}
\bibitem{noi3}
We checked several options, based on calculations
for Be-H and for isolated H in GaN \cite{mrs}, including 
H bound at N vacancies.
See also Ref. \cite{neuprl}
\end{references}
\end{document}